# Investigating the Prominence and Severity of Bugs and Glitches Within Games and Their Effects on Players' Experience

Jessica Backus
Purdue University, Computer Graphics Technology
backus@purdue.edu

**Abstract**
Different errors that occur in video games are often referred to as "glitches" or "bugs." The goal of this exploratory research is to understand how these glitches and bugs within video games affect a player's experience. To do this, I reviewed relevant literature and performed observations of these different errors in different games via Twitch livestreams. I then performed thematic analysis with the observation data and generated themes that tie back into to the relevant literature. Most of the current literature focuses on the "what" and "how" behind bugs in games, but very little on the implications of these bugs on the overall experience for the players, and what patterns of behavior may emerge because of them.

**Keywords:** Video Games, Self-determination Theory, Glitches, Bugs, Errors, Streaming, Cheating, Bug Exploiting.

**1 Introduction**
Like with any other software, video games are prone to errors. These appear in the form of bugs and glitches, and there are many types of each. Due to the complexity of these errors, people often confuse the two, or use the terms interchangeably. To clarify the meaning behind these terms used throughout this paper, a "bug" is an error that occurs within a video games software "behind the scenes." It can cause errors in the game's systems and how it functions. Whereas a "glitch" is an error that can occur when the game's systems are all performing as intended. For example, if a player cannot interact with a game object, this is a bug. But if a game object appears to "clip" through another object, then this is a glitch. Because a video game's success is dependent on a positive player experience, it's important to understand how players perceive these different errors and how they affect the player's overall experience. To do this, I ask the following questions:

*1.1 Research Questions*
1. What is the threshold of bugs that make a game "acceptable" vs. "unplayable?"
2. What makes a bug "severe?"
3. "What are the most common types of bugs?"

The first research question will allow me to grasp what players deem as "acceptable" in a game, as far as how much a bug (or bugs) impacts their choice to play a specific game. This question is largely subjective, and therefore will rely more on the observation insights. The second question examines how severe bugs are perceived and determined. This question has subjective and objective elements, therefore observations and literature review can both provide insights for this question. With the many different types of bugs (that will be covered by research question #3), it is important to know which bugs are considered more severe by the players and affects their experience the most, not just what different development teams deem "severe" by their own definitions. Lastly, the third question is purely objective in nature, so I will use the information found in the literature to inform me of the different bugs that I observe in my observations.

## 2 Methods

To investigate the research area, a literature review was performed to gauge the current efforts to understanding this phenomenon. After gathering many resources, a method sometimes called "netnography" [source] was used to observe gaming livestreams; one on the livestreaming platform Twitch and the other on YouTube (that was a video posted of a recording of a Twitch livestream). Both observations were not performed during the livestream simply due to the nature of glitches; it is very hard to catch them live because of how unexpected they can be (unless done intentionally, which will be discussed in later sections). One of the recordings also included the audience chat and their responses were also recorded where relevant. After obtaining a set of thick field notes from these observations, they were coded and analyzed according to thematic analysis methods.

## 3 Overview of Relevant Literature

### *Glitch or a Feature?*

One study by Švelch [6] investigates the use of glitches in games with Microtransactions. Specifically, they gathered qualitative data from developers, fans, and the press regarding glitches within two of EA's games: *Dead Space 3* and *Mass Effect 3*. The significance of the presence of microtransactions in these games with glitches means that players were able to obtain in-game benefits without purchasing microtransactions by utilizing certain glitches. While its easily argued that these glitches give players an unfair advantage, the fact is that these microtransactions also gave players these same advantages- just not for free. In the end, the only one harmed in this scenario is EA. However, for one game/glitch they swiftly deemed this not just a glitch, but a cheat. The other, they deemed it a feature. Why? The first glitch gave players an *immediate* advantage whereas the second one still required time and effort to get an advantage (but still less time/effort than intended). Because the advantage is not instant, their microtransactions (that provide instant upgrades) was still considered useful.

The first glitch was made possible when any weapon in the game could be used to shoot rockets (the strongest weapon in the game). Fans stated that "missile glitching just plain breaks the game" [6]. Because of how drastically this glitch affected player's experience and how different it was from the developer's intentions for those weapons, everyone was quick to agree that this was indeed a glitch and not a feature.

### *Bug Exploitation = Cheating?*

A popular topic regarding bugs is the exploitation of them. A study by Lewis et. al looked at different player behaviors in a popular multiplayer game, Counterstrike: Global-Offensive in the context of Esports [4]. Bugs within this context are highly debated. Some argue that anything that gives an advantage over another player is a form of cheating [1]. However, others argue that they are simply utilizing every aspect of the game and are not altering the game in anyway, therefore it is not cheating. In esports tournaments, this debate becomes even more intense due to the real-life stakes (often monetary rewards). Some tournaments must explicitly state that some bugs/glitches are not allowed, whereas some have a "gentleman's agreement" to not exploit known bugs [4]. In the case that a team does exploit bugs for an advantage, they are either disqualified, they forfeit, or nothing can be done against them at the time if it was not written in the rules. However, this often leads to fan backlash, so there are still negative repercussions. Much like "normal" sports, cheating (and bug exploiting) is constantly under review.

### *Creating a Bug Taxonomy*

A paper from Lewis et. al [4] created a taxonomy of the different types of bugs that they found throughout various YouTube videos that users uploaded displaying the bugs. They claim that the sheer amount of videos like this on YouTube provides a wealth of examples, far more than a single study could ever collect. They then categorize these bugs they found, describing the bugs and what games they were found in. They also break down the different bugs into their faults, errors, and failures. This method identifies what exactly is going on in the games systems as well as what outcomes they create and what the user sees. By using this method, they can help game developers better understand not only exactly how the bugs occur, but how they can identify the exact approach to fixing/preventing them.

*3.1 Types of Bugs*

Because the purpose of Lewis's [4] paper is to provide information to developers in the pursuit of correcting bugs, the taxonomy they create uses the vernacular used in the gaming industry (ex: "Invalid value change", "Object out of bounds for any state"). For the purpose of this paper, I used their taxonomy to improve my own understanding of bugs and I then "summarized" their information by grouping some of their bug types together. Below in Figure 1 is a chart of this summarization and some examples of each bug type. The method of grouping these bug types center on the player, objects, and actions. There is also a 4$^{th}$ category that focuses on Lagging, because this phenomenon affects how the game itself is running and can create several different outcomes (visual issues, incorrect actions, invalid player position, etc.).

## 4 Observations

To learn about the subjective views that players have about glitches and bugs, I took on a similar approach to Lewis et. al [4], but instead of simply viewing shorter (and often edited) YouTube videos, I watched recordings of livestreams from Twitch. One of these livestreams included the audience chat and as the glitches or bugs occurred, I noted their live reactions. This is included because having an entire group of

**Types of Glitches**

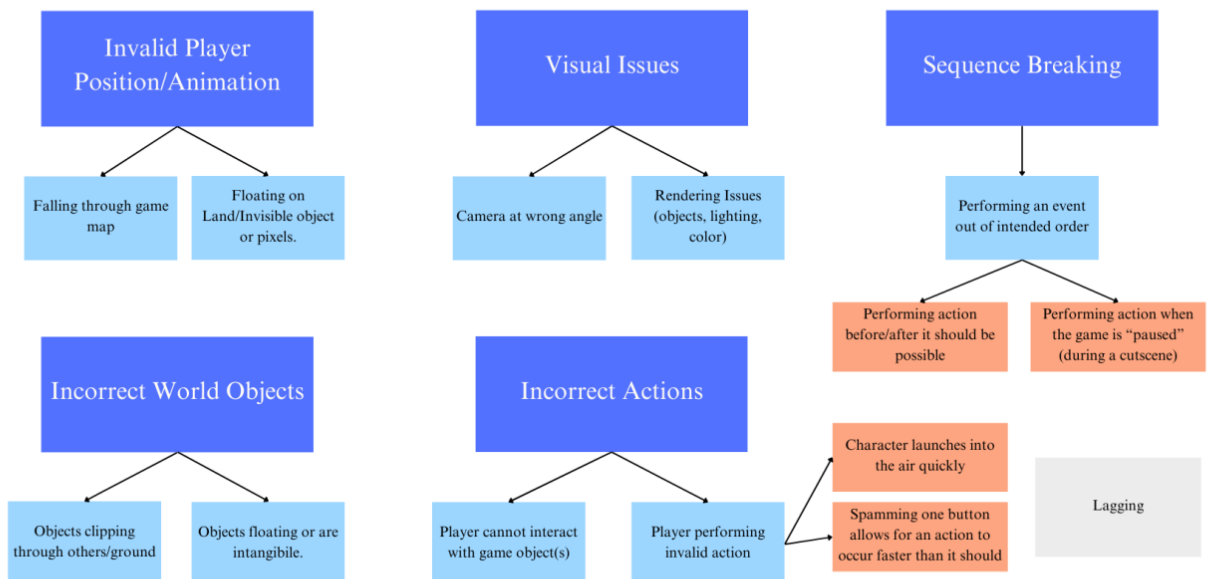

Figure 1: Types of Bugs generated from the taxonomy discussed in [4].

people's reactions can provide further insights outside of just the streamer's reactions and can be useful in gauging how these bugs and glitches are interpreted. Irwin [3] mentions that in certain situations, this method can be helpful, especially in highly subjective/debatable contexts (such as cheating). This can also provide a look at the specific online community's culture as well. The two games that were played in the observations were *Call of Duty: Warzone (2021)* and *The Long Drive* [7, 8]. Both games were multiplayer, so the streamer played with their friends (rather than random others). Both observations were of the same streamer, but they played with different friends each time. Each observation was 30 minutes, totaling an hour.

*4.1 Analysis*
The analysis of the field notes taken during observations started with coding using the web-based software Dovetail. Some of the codes were focused on the topic discussed in that moment ("glitch type", "multiplayer game") whereas some were more abstract ("reactions", "outcome of bugs"). Once the observations were coded, I then looked for patterns in the observations to generate themes from. Below in Figure 2 is a section of the canvas used to analyze the data/codes as an example of my process.

**5 Results and Discussion**

The first and easiest observations to take were the purely objective ones including the answer for my third research question: "What are the most common types of bugs?" In both observations, there were examples of the player or an object "clipping" through another (with different end results), which falls under "Invalid Player Position" and the "Incorrect World Objects" categories I discuss in the "Types of Bugs" section. Hind and Bell [2] also state that "[Clipping] happens in virtually all games […] it can happen even in a game with great art direction and high visual polish…" So the fact that this was the most common bug I observed shouldn't come as a surprise. The second most common observation was "Incorrect Actions," often in the form of the player not being able to interact with game objects as intended. What was very interesting was that this bug occurred after a glitch occurred, in both scenarios. So not only does this imply that glitches can lead to bugs, but bugs can happen after a glitch occurred intentionally or unintentionally.

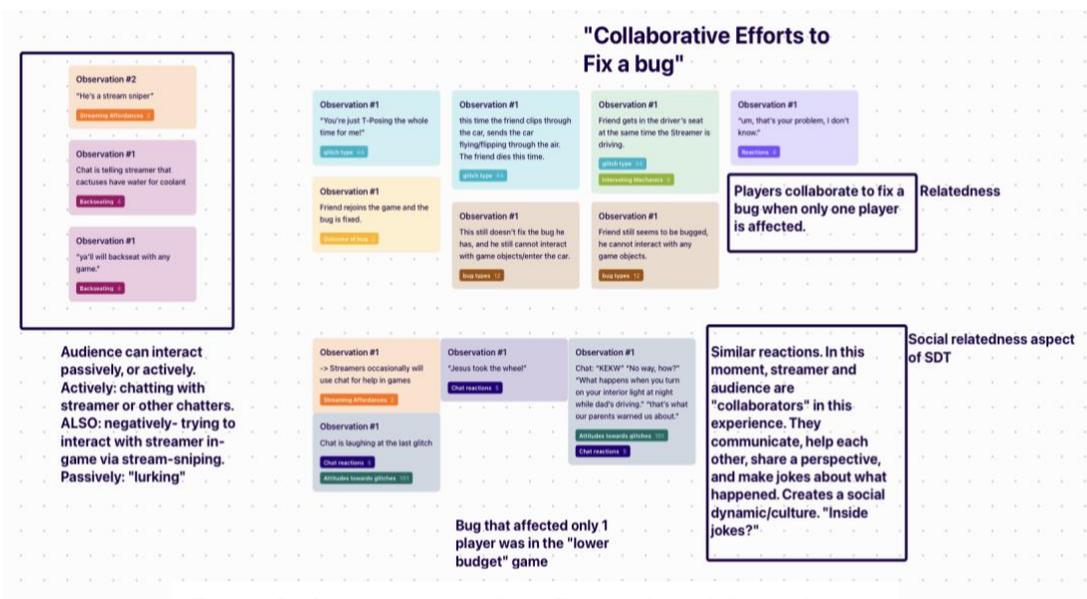

Figure 2: Canvas snippet from Dovetail used for analysis.

The big difference between both observations is that a glitch occurred in both, but one was triggered accidentally whereas the other was triggered intentionally. The accidental glitch happened in *The Long* Drive when a player tried to enter a car in the game, and they ended up clipping through it, and the car started clipping through the ground. This killed the player and when they respawned, they could not interact with any game object. The intentional glitch occurred in *CoD: Warzone* when one of the streamer's friends heard about a glitch online where you could place an ammunitions crate on a truck and then place a sentry turret on the ammunitions crate, essentially creating a mobile sentry turret. They tried to do this, and while they were able to successfully place the turret on a truck, it caused the truck to start clipping through the ground and it became nearly impossible to drive it normally, and the streamer even clipped through the truck. The driver of the truck also died like in the first case but did not respawn due to the game's rules. The implications of this are that bugs can be triggered by players when trying to exploit certain glitches, or they can happen accidentally through normal gameplay.

Another key difference in these situations is who was affected by the bugs. In *The Long Drive*, the glitch happened to only one player and the subsequent bug that occurred also only affected that player. In *Warzone*, the glitch happened to all players and the bug of the truck clipping also affected all players. So, another takeaway is that bugs can affect one or multiple players. If a bug occurs for all players, it can be argued that this bug is more severe. The theme that I generated from this is "Who the bug affects." Another bug that was observed in *Warzone* was when entering certain buildings, a sun/lens flare was visible through the roof, creating an unpleasant glare on the player's screen. The streamer and his friends all experienced this bug and even commented their distaste of its presence, "*Ow! It's brighter than it is normally! Awful.*" The nature of this bug was that it affected one of the games basic components: the visuals. If a bug affects some of the basic systems of the game such as the visuals, sound, or controls, it can also be argued that these bugs are more severe, not only because they are easier to measure, but also because they are universally agreed to create a negative experience. Like with the missile glitch discussed by Švelch [6], bugs like this that clearly go outside of the developer's intention for the game are quickly viewed as bugs. Hind and Bell [2] also discuss how some bugs can be purely objective whereas some are subjective. It's clear to see that bugs that affect basic components are more objective, and therefore more severe. This led to my theme of "Bugs affecting Basic Functions."

The first game played, *The Long Drive*, was made by a smaller company and had a smaller budget compared to the game giant Activision and their *Call of Duty: Warzone* title. Therefore, it's unsurprising that this game had more bugs than the latter. However, these bugs did not make the game "unplayable." In fact, the streamer, his friend, and the audience seemed to rather enjoy the bugs that occurred in the game. Only at one point did the game come close to the "unplayable" threshold, where one of the players had to leave the game and rejoin in attempts to fix the bug he was experiencing. Luckily for him, this worked, and he was able to continue playing. The bug he experienced was the one mentioned before, where he was unable to interact with any game object. It can be argued that interaction is also a basic component of any game and therefore, this bug can be considered a severe bug that hindered gameplay. So, to answer the first research question, it's not always about the quantity of bugs in a game that make it close to unplayable, but rather how severe they are and how severely they impact the players ability to accomplish basic tasks.

## 6 Limitations

As mentioned previously, I performed 1 hour total of observations, both with the same streamer, just with different games. While 1 hour helped provide me with interesting insights, I could have more with more observations, perhaps with other streamers and other games. Also, it would be ideal to observe glitches live but due to their unpredictable behavior, it is difficult to observe unintentional glitches this way. When it comes to observing intentional glitches, there are few people that know how to successfully perform an intentional glitch, and these are often done in specific contexts, such as "Speed-running" in single-player games, rather than in multiplayer games, since oftentimes these glitches are patched, or the players are banned.

It's important to note that I am a member of the online communities that I observed. While I wasn't an active participant at the time of observation, the fact that I participate in this community in general may lead to some bias in my conclusions and my data collection. However, it can also be argued that I can provide "inside knowledge" about certain observations, giving some extra context that may not otherwise be there for non-members.

## 7 Implications and Conclusions

Understanding how players perceive different bugs and glitches within games can help inform the development team which bugs/glitches they should focus their efforts on fixing. This can lead to saved time, money, effort, and most importantly, better the player's experience. Efforts of which bugs to focus on first (or at all) are already being done by game testers [2], but these are often guided by past or current design methods. They do take into consideration the "average" player (instead of diehard fans that will nitpick every detail) but learning about specific cases can also provide useful insights. Players that purposefully glitch a game, whether it be for game mastery (fulfilling the competency need in Self-determination Theory [5]), potential funny moments, or exploring the game's mechanics, they all represent different types of players and motivations for playing. If developers understand why different player types enjoy certain games and the bugs within those games, then they can actively choose which bugs to patch and which they can deem a feature.

*Studies Conference Brno 2014*. Central and Eastern European Game Studies Conference, Brno, Czech Republic. https://books.google.com/books?hl=en&lr=&id=gnfkDwAAQBAJ&oi=fnd&pg=PA55&dq=Negotiating+the+Glitch+Identifying+and+Using+Glitches+in+Video+Games+with+Microtransactions&ots=hxQxn__mQW&sig=-nUaxN-7QYr4gaKl635QKtHhosw#v=onepage&q=Negotiating%20the%20Glitch%20Identifying%20and%20Using%20Glitches%20in%20Video%20Games%20with%20Microtransactions&f=false

### *8.1 Sources of Observation Livestreams*

[7] SMii77Y (2023). *going on a long drive with kryoz*. [Video]. Twitch. **https://www.twitch.tv/videos/1976153422**

[8] SMii7Y (2021). *[SMii7Y VOD] The Most Chaotic Warzone Video on YouTube*. [Video]. YouTube. https://www.youtube.com/watch?v=8Gb-2edjhnI&list=PLTQcm9IcdydLoyKMSRbUuPocsgGXQfq0Q&index=34